\documentclass[10pt,journal,cspaper,compsoc]{IEEEtran}

\usepackage{expdlist}
\usepackage{color}
\usepackage{enumerate}
\usepackage{graphicx}

\ifCLASSOPTIONcompsoc
\else
\fi
\ifCLASSINFOpdf
\else
\fi
\begin{document}
\title{Fighting network space: it is time for an SQL-type language to filter phylogenetic networks}

\author{Steven Kelk,
        Simone Linz,
        and~David A. Morrison% <-this % stops a space
\IEEEcompsocitemizethanks{\IEEEcompsocthanksitem S. Kelk is with the Department of Knowledge Engineering (DKE), Maastricht University, P.O. Box 616, 6200 MD Maastricht, The Netherlands. Email: steven.kelk@maastrichtuniversity.nl\protect
\IEEEcompsocthanksitem S. Linz is with the Center for Bioinformatics, University of T\"ubingen, 72076 T\"ubingen, Germany, and the Department of Mathematics and Statistics, University of Canterbury, Christchurch, New Zealand. Email: linz@informatik.uni-tuebingen.de.
\IEEEcompsocthanksitem D. A. Morrison is with the Section for Parasitology, Department of Biomedical Sciences and
Veterinary Public Health, Swedish University of Agricultural Sciences, 751 89 Uppsala, Sweden. Email: david.morrison@slu.se.
}% <-this % stops a space
\thanks{}}

\IEEEcompsoctitleabstractindextext{%
\begin{abstract}
%\boldmath
The search space of rooted phylogenetic trees is vast and a major research focus of recent decades has been the development of algorithms to effectively navigate this space. However this space is tiny when compared with the space of rooted phylogenetic networks, and navigating this enlarged space
remains a poorly understood problem. This, and the difficulty of biologically interpreting such networks, obstructs
adoption of networks as tools for modelling reticulation. Here, we argue that the superimposition of biologically motivated constraints, via an SQL-style language, can both stimulate use of network software by biologists and potentially significantly prune
the search space.
\end{abstract}
% not want either math or citations to appear in the abstract.

% Note that keywords are not normally used for peer review papers.
}

\maketitle

% is not selected <OR> if conference mode is selected - because compsoc
% conference papers position the abstract like regular (non-compsoc)
\IEEEdisplaynotcompsoctitleabstractindextext

% \fi
\IEEEpeerreviewmaketitle

\section{Introduction}
% Computer Society journal papers do something a tad strange with the very
% first section heading (almost always called "Introduction"). They place it
% ABOVE the main text! IEEEtran.cls currently does not do this for you.
% However, You can achieve this effect by making LaTeX jump through some
% hoops via something like:
%
%\ifCLASSOPTIONcompsoc
%  \noindent\raisebox{2\baselineskip}[0pt][0pt]%
%  {\parbox{\columnwidth}{\section{Introduction}\label{sec:introduction}%
%  \global\everypar=\everypar}}%
%  \vspace{-1\baselineskip}\vspace{-\parskip}\par
%\else
%  \section{Introduction}\label{sec:introduction}\par
%\fi
%
% Admittedly, this is a hack and may well be fragile, but seems to do the
% trick for me. Note the need to keep any \label that may be used right
% after \section in the above as the hack puts \section within a raised box.

\IEEEPARstart{R}{ooted} 
phylogenetic networks are extensions of rooted phylogenetic trees to explicitly incorporate reticulation
events such as lateral gene transfer and hybridization, modelled as nodes with two or
more parents (e.g. the network shown in Figure~\ref{example} has a reticulation $r_1$ with the three parents $p_1$, $p_2$, and $p_3$). While the potential of such networks for hypothesis generation and testing is increasingly recognised, a number of obstacles prevent their widespread use by evolutionary biologists. First, the space of rooted phylogenetic networks is vast, far larger than the space of rooted trees, and even heuristically navigating this space is a formidable
computational challenge. Second, hypothesis-testing techniques that are standard in the tree literature, such as the ability to query whether there is support for a particular clade, 
%being monophyletic, 
are not yet well-developed. These two
problems often coincide in the sense that, depending on the specific context, a large number of networks in the
search space will be biologically irrelevant. For this reason it is both biologically and computationally attractive
that biologists should be able to describe \emph{a priori}, via a user-friendly SQL-style (Structured Query Language) modelling language, those networks which should (or should not) be taken into consideration. Such constraints can help biologists interpret the output of network-building software and, when incorporated into the search algorithms used by such software, potentially allow the search
space to be dynamically pruned. Furthermore, they can equally be used \emph{a posteriori} to filter an already given set of candidate networks for biological relevance. 
Historically, the idea of using constraints to reduce the search space of phylogenetic trees dates back to at least~\cite{Sankoff:1982ti}, who pointed out that a complete search of a smaller tree space could be better than a heuristic search of a larger space, in terms of finding the optimal tree. Inspired by this idea, we propose an outline for such a constraint-based framework for phylogenetic networks.

\section{SQL-style network modelling and what we can learn from trees}

\begin{figure*}[!t]
\centering
\includegraphics[width=15cm]{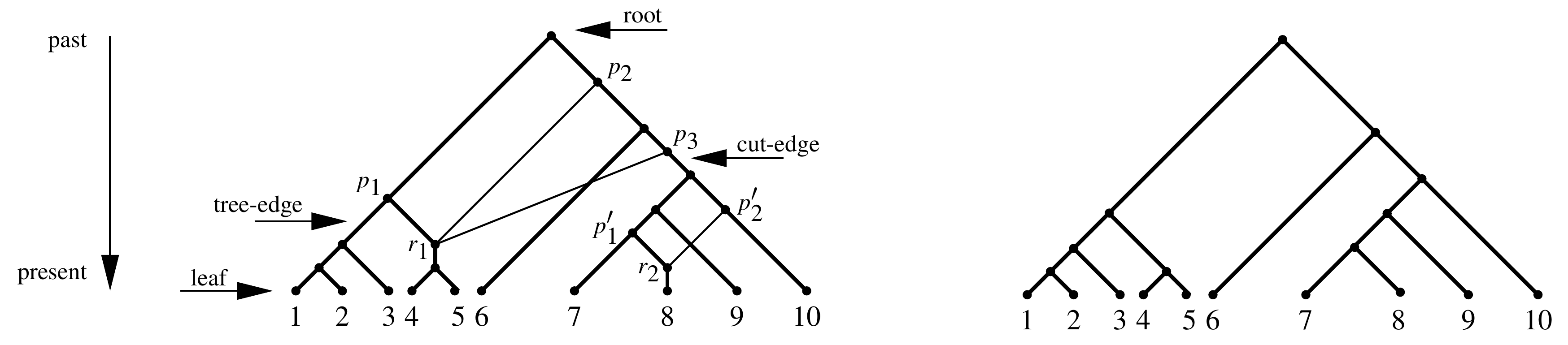}
\caption{A network $N$ (left) and a rooted phylogenetic tree (right) that is embedded in $N$ as indicated by the thicker edges in $N$. All edges are directed downwards. Note that $N$ has two reticulations $r_1$ and $r_2$.}
\label{example}
\end{figure*}

As indicated in Figure~\ref{example}, a rooted phylogenetic network, henceforth simply \emph{network}, is an extension of the rooted
phylogenetic tree to the space of rooted directed acyclic graphs. For a technical description
of their characteristics we refer the reader to \cite{HusonRuppScornavacca10}. Networks are often constructed
as parsimonious summaries of incongruence within a set of trees. A common goal is to construct a network that is
as parsimonious as possible, in terms of the number of reticulations, and which has all the input trees simultaneously embedded within it (e.g. the tree that is shown on the left-hand side of Figure~\ref{example} is embedded in the network that is shown on the right-hand side of the figure). Methods that work directly on sequence data are also emerging and show considerable promise, as well as approaches that model duplication, loss, transfer, and incomplete lineage sorting events by reconciling a gene tree with a species tree (e.g.~\cite{Stolzer:2012dq}). However, severe computational intractability (NP-hardness or worse) is a recurring feature of almost all explicit network methodologies, greatly limiting the scope of their application. The core of the problem is that, even for a small number of taxa and reticulations, the space of networks is vast, and even heuristic traversal of this space is problematic. One way of trimming the space of networks is to heavily constrain the
number of reticulations and/or their relative location in the network. Although beneficial from a tractability perspective such constraints should be first and foremost biologically well-motivated. Indeed, constraint-based pruning is a feature
that one often encounters in tree-building software to test hypotheses (e.g. the SOWH test~\cite{Goldman:2000vx}). Software such as  PAUP*~\cite{swofford02} and RAxML~\cite{stamatakis2005raxml}, for example, allow the user to restrict the search of tree-space to trees that contain a certain clade or that are consistent
with a given tree backbone (i.e. a constraint tree). Such restrictions allow the user to test support for competing
clade hypotheses, an experimental technique that is used extensively in practice.

As Figure~\ref{example} suggests, networks have features that cannot be described simply in terms of clades.
%Some features are biologically highly questionable, such as allowing lateral transfers between taxa that cannot possibly  %have
%co-existed in time. Nevertheless, these features persist, and a major cultural reason for this is that algorithm developers %are hesitant to \emph{a priori} exclude candidate networks from the space of solutions that might be biologically %meaningful,
%a subjective decision that they invariably do not feel qualified to take. 
It is unlikely, and unreasonable to expect, that biologists will reach a single, unified consensus on which network
features are meaningful and which are not. However,
in a given experimental context, and guided by the
data at hand, a biologist often already has some insight into which networks do, and do not, constitute plausible
hypotheses. The challenge therefore is to provide biologists with an easy-to-use tool that allows them to formally articulate
these insights. For maximum flexibility, such a tool should allow both certain natural atomic constraints and SQL-style compound constraints. 
We note here that SQL was originally developed for managing and retrieving database content by using complex queries, but the concept is now used widely in computational science.
The atomic constraints should allow fundamental characteristics of the
candidate network to be tested (and, ideally, should themselves be computationally tractable). Some examples include: 
\begin{description}[\compact]
\item[(a)] A given subset of taxa must be below a \emph{cut-edge}, i.e. a locally isolated part of the network.
\item[(b)] A given subset of taxa must be below a \emph{tree-edge}, i.e. a purely tree-like part of the network.
\item[(c)] A given taxon $x$ should be a hybrid of two other designated taxa $y$ and $z$.
\item[(d)] A given tree should be embedded in the network.
\end{description}
To illustrate, taxa 7, 8, 9, and 10 are below a cut-edge and taxa 1, 2, and 3 are below a tree-edge in the network shown in Figure~\ref{example}. Furthermore, taxon $r_2$ is a hybrid of the two taxa $p_1'$ and $p_2'$.

In addition to atomic constraints that describe certain topological characteristics of a network, one could also include statistically motivated constraints. For example, one may wish to consider only those networks whose probability of a given gene tree topology exceeds some user-defined value (for details, see \cite{Yu:2012hx}). Note, also, that constraints can be either positive (specifying characteristics that must appear in the final network) or negative (forbidding certain characteristics).

Lastly, as mentioned above, atomic constraints can be used as building blocks to design more powerful compound constraints. 
The next example combines three atomic constraints, and might be useful if one has more detailed information about the evolutionary history of a subset of the taxa under consideration. In such a case, one could build the following SQL-type query:\\

\noindent  \textbf{SELECT} those networks whose number of reticulations is below a certain threshold \textbf{AND} that have a given subset of the taxa below a cut-edge \textbf{WHERE} a time-consistent labeling can be assigned to the nodes of the subnetwork below the cut-edge. \\
 
\noindent  Here, a time-consistent labeling is a labeling on the nodes of a (sub)network such that reticulation events occur only among contemporaneously existing taxa. For example, the network shown in Figure~\ref{example} is not time-consistent because the two parents $p_2$ and $p_3$ of $r_1$ cannot have the same timestamp, whereas the subnetwork below the indicated cut-edge in the same figure is indeed time-consistent.

%\section*{\blue{The biological relevance of SQL-style constraints}}
\section{Constraints for filtering and pruning}
The purpose of SQL-style constraints in the construction and analysis of networks is twofold. First, they can be used {\it a posteriori} to filter a given set of candidate networks resulting from an analysis that reconstructs networks from a data set (e.g. characters, trees, clusters). Since many of these methods are based on combinatorial frameworks, the set of optimal candidate solutions can be quite large. For example, an analysis of a well-known grass data set that finds all networks containing two given trees, each on 40 taxa, and whose number of reticulations is minimized, results in a set of (at least) 2268 optimal solutions. Obviously, validating all optimal networks by hand becomes a tedious task. In order to support the biologist in this part of an analysis, \cite{huson13} describes two constraints that are available as part of the Dendroscope program~\cite{huson2012dendroscope} and can be used to filter or rank a list of previously generated networks. Second, SQL-type constraints can also be defined  before any analysis so that the search space of networks, which is vast, can be pruned dynamically. Since the space of networks is, in general, infinite for a fixed number of taxa, even a small number of constraints can greatly aid the computational  process and significantly reduce its running time, so that instances of a larger input size can potentially be solved exactly. In the context of phylogenetic trees, \cite{Constantinescu:1986ij} showed that the use of a constrained tree 
 reduces the search space remarkably by calculating the difference between the number of trees of a fixed size that need to be considered in an unconstrained search and the number of trees of the same size that are compatible with a given constraint tree.
 
 The technicalities
involved in dynamic pruning will be nontrivial, but we draw inspiration from a number of
``branch and bound''-style pruning techniques that are already being used, albeit in an ad hoc fashion, in the phylogenetic network literature. For example, algorithms that construct
networks that contain embeddings of triplets (or clusters) will cease to explore a branch of
the network search space if the partially constructed network already fails to contain a certain triplet, because adding more taxa to the network will never recover the missing triplet \cite{cass,simplicityAlgorithmica}. Adding additional
constraints should allow for even more aggressive pruning of the network search space.
The step from high-level constraints to low-level pruning is a topic we hope to return to in a
forthcoming article.

Furthermore, and perhaps most fundamentally, by using SQL-type constraints, biologists have more control on the output of programs that reconstruct networks. In fact, they can go beyond the widespread concept of regarding an algorithm as a black box, and actively engage in the construction of networks by adding extra biological information to it and, therefore, reduce the risk of misanalyses.

\section{Conclusion}

The space of phylogenetic networks is huge, and this is an obstacle from both a computational
and interpretational viewpoint. We propose the development of an SQL-style constraint-based
language that will allow the imposition of biologically relevant constraints on this space,
thus enhancing the utility of phylogenetic networks for biologists, and potentially cutting down the
search space of phylogenetic networks to a more reasonable size.

% or


\ifCLASSOPTIONcompsoc
  \section*{Acknowledgments}
\else
  \section*{Acknowledgment}
\fi

S.L. was supported by a Marie Curie International Outgoing Fellowship within the $7^{\textnormal{th}}$ European Community Framework Programme.

\ifCLASSOPTIONcaptionsoff
  \newpage
\fi



%\bibliographystyle{IEEEtran}

\bibliographystyle{IEEEtran}
\bibliography{IEEEabrv,networksSQL}

\end{document}